\documentclass[aps,pre,showpacs,twocolumn,groupedaddress,superscriptaddress]{revtex4}
\usepackage{graphicx}%
\usepackage{amsmath}
\usepackage{dcolumn}

\begin{document}

\title{Asymmetry--induced effects in coupled phase oscillator
ensembles:\\
Routes to synchronization}

\author{Jane H.~Sheeba}%
\affiliation{Department of Physics, Lancaster University,
Lancaster, LA1 4YB, UK}
\affiliation{Centre for Nonlinear Dynamics, School of Physics,
Bharathidasan University, Tiruchirappalli - 620 024, Tamilnadu, India}

\author{V.~K.~Chandrasekar}%
\affiliation{Centre for Nonlinear Dynamics, School of Physics,
Bharathidasan University, Tiruchirappalli - 620 024, Tamilnadu, India}
\affiliation{Department of Physics, Lancaster University,
Lancaster, LA1 4YB, UK}

\author{Aneta~Stefanovska}%
\affiliation{Department of Physics, Lancaster University,
Lancaster, LA1 4YB, UK}
\affiliation{Faculty of Electrical Engineering, University of Ljubljana, Tr$\breve{z}$a$\breve{s}$ka
25, 1000 Ljubljana, Slovenia}

\author{Peter~V.~E.~McClintock}%
\affiliation{Department of Physics, Lancaster University,
Lancaster, LA1 4YB, UK}

\date{\today}

\begin{abstract}
A system of two coupled ensembles of phase oscillators can
follow different routes to inter-ensemble synchronization.
Following a short report of our preliminary results [Phys. Rev.
E. {\bf 78}, 025201(R) (2008)], we present a more detailed
study of the effects of coupling, noise and phase asymmetries
in coupled phase oscillator ensembles. We identify five
distinct synchronization regions, and new routes to
synchronization that are characteristic of the coupling
asymmetry. We show that noise asymmetry induces effects similar
to that of coupling asymmetry when the latter is absent. We
also find that phase asymmetry controls the probability of
occurrence of particular routes to synchronization. Our results
suggest that asymmetry plays a crucial role in controlling
synchronization within and between oscillator ensembles, and
hence that its consideration is vital for modeling real life
problems.
\end{abstract}



\maketitle

\section{Introduction}
Ensembles of coupled oscillators are ubiquitous in nature. They
arise in diverse areas of science including physics, biology,
chemistry, neuroscience, social, electrical and ecological
systems. Examples include synchronous emission of light pulses
by populations of fireflies \cite{Buck:68}, synchronized firing
of cardiac pacemaker cells \cite{Peskin:75}, synchronization in
ensembles of electrochemical oscillators
\cite{Kiss:02a,Kiss:02b}, both short- and long-range
synchronization in the brain (within and between neuronal
ensembles) \cite{Hansel:92,Bressloff:99,Golomb:01}, emission of
chirps by a population of crickets \cite{Walker:69}, and
synchronous clapping of audiences in auditoria. Research into
the dynamical properties of large ensembles of this kind has
been a subject of intense interest since the 1960s
\cite{Winfree:67,Kuramoto:84,
Strogatz:00,Strogatz:01,Pikovsky:01}. Mean-field theory
facilitates the study of such ensembles by reducing the
dynamics of a number of oscillators to the dynamics of their
mean field, i.e.\ effectively of a single oscillator. In
principle, each oscillator in the ensemble contributes to the
dynamics of the mean field, so that the collective dynamics of
the entire ensemble can be represented by the dynamics of the
mean field. This approach has a good analytical background that
enables identification of bifurcation boundaries and stability
criteria for understanding the synchronization dynamics of the
ensemble. Although the mean field approach suggests
consideration of the dynamics of just one oscillator in place
of the ensemble dynamics, recent research has identified new
phenomena such as intra-ensemble and inter-ensemble clustering
\cite{Jane:08a} that can only be understood in terms of
ensembles. Thus one should expect to model natural systems
comprised of interacting entities as ensembles of coupled
oscillators, rather than always approximating them as a single
oscillator.

Synchronization, or concurrence between oscillatory systems, is
a remarkable phenomenon that is often inescapable for coupled
oscillators. Phase synchronization was first reported by the
Dutch physicist Chistiaan Huygens well back in the 17th century
based on his observation of two pendulum clocks that persisted
in precise antiphase, seemingly indefinitely. Thereafter, the
phenomenon of synchronization has been studied theoretically
\cite{Winfree:80,Kuramoto:84,Pikovsky:96,Strogatz:00,Rulkov:01}
and experimentally
\cite{Hansel:92,Haken:93,Bressloff:99,Neda:00,
Golomb:01,Topaj:01,Kiss:02a,Kiss:02b} in great detail. It is
well known that the control of synchronization in natural
systems \cite{Blasius:03,Montbrio:03,Blasius:05} is of great
important. The occurrence of synchronization is very important
for e.g.\ lasers and Josephson-Junction arrays
\cite{Trees:05,Rogister:07}, cardio-respiratory synchronization
\cite{Schafer:98,Lotric:00} or temporal coding and cognition
via brain waves
\cite{Singer:99a,Singer:99b,Fries:05,Yamaguchi:07,Jane:08b}.
However the emergence of synchronized oscillations can also
give rise to undesirable effects, as in the case of epileptic
seizures \cite{Goldberg:02,Timmermann:03}, Parkinson's tremor
\cite{Percha:05,Zucconi:05}, or pedestrians on the Millennium
Bridge \cite{Strogatz:01}.

In real systems, the interactions between the oscillators are
often asymmetric. Examples include cardio-respiratory
\cite{Stefanovska:99,Palus:03} and cardio-$\delta$ (EEG)
interactions \cite{Stefanovska:07}, interactions among
activator-inhibitor systems
\cite{Daido:04,Daido:06,Daido:07,Kiss:02b}, coupled circadian
oscillators \cite{Fukuda:07}, and the interactions between
ensembles of oscillators in neuronal dynamics
\cite{Sherman:92,Roelfsema:97,Singer:99a}. Neglecting coupling
asymmetry, i.e.\ assuming symmetric interactions, is an
approximation that may simplify the analysis but which may also
lead to a model that fails to describe important phenomena
occurring in the system. We have already reported
\cite{Jane:08a} novel global clustering phenomena, and novel
routes to inter-ensemble synchronization that occur only in the
case of asymmetrically interacting systems. It is evident,
therefore, that explicit consideration of asymmetry in the
interaction may be essential to create a realistic model.

In this paper, we supplement the preliminary account
\cite{Jane:08a} of our investigations of two asymmetrically
interacting ensembles of oscillators by providing additional
detail of the different synchronization regimes, and we extend
it by reporting the effects induced by noise asymmetry. We
thereby emphasize the importance of asymmetry -- in coupling,
noise and phase -- in such systems. We show that it is the
coupling and phase asymmetries that control their
synchronization. We also report the occurrence of certain novel
routes to inter-ensemble synchronization. We show that these
routes are characteristic of asymmetrically interacting
ensembles of oscillators and that they cannot occur in systems
where the interactions are symmetrical. These results yield new
insights into how synchronization arises in coupled oscillator
ensembles. This understanding is an essential prerequisite for
the development of control schemes, paving the way to possible
ways of controlling synchronization in real systems.

We introduce the model of asymmetrically interacting ensembles
of oscillators, and define their mean field, in Sec.\
\ref{Model}. In Sec.\ \ref{Stability} we discuss analytically
the stability of the {\it incoherent} (i.e.\ unsynchronized)
state in the thermodynamic limit and consider how it can be
modelled numerically. Sec.\ \ref{SychRegimes} defines the five
distinct synchronization regimes that we have identified, and
discusses in turn how each of them is influenced by asymmetry
in coupling, noise, and phase. The several routes followed to
synchronization, and between different synchronization regimes,
are discussed in Sec.\ \ref{routes}. Finally, in Sec.\
\ref{summary} we summarize the main results and draw
conclusions.

\section{Coupled phase oscillator ensembles}
\label{Model} The energy emitted or absorbed
by an individual oscillator in the ensemble
will alter the physical states of the neighbors to which it is
coupled; in particular, the periods of its neighbors are
altered (either lengthened or shortened). The way in which the
period is altered depends on the state of the neighbouring
oscillator at the moment when it receives the impulse. One of
the commonest scenarios to consider is an ensemble of nonlinear
oscillators evolving in a globally attracting limit cycle of
constant amplitude. Such oscillators are called limit cycle or
phase oscillators. If they are coupled in such a way that they
will not be perturbed sufficiently to leave their limit cycles,
then one degree of freedom is enough to describe the system
dynamics. Let us consider a system of two asymmetrically
interacting ensembles of oscillators (AIEOs). Their phase
dynamical equations can be written as \cite{Kuramoto:84}
\begin{eqnarray}
\label{mod01}
 &&\dot{\theta_i}^{(1,2)}= \omega_i^{(1,2)} - \frac{A^{(1,2)}}{N^{(1,2)}}
 \sum_{j=1}^{N^{(1,2)}}
f(\theta_i^{(1,2)}-\theta_j^{(1,2)}+\alpha^{(1,2)}) \nonumber \\
 &&-
 \frac{B}{N^{(2,1)}}\sum_{j=1}^{N^{(2,1)}}
 h(\theta_i^{(1,2)}-\theta_j^{(2,1)}+\alpha^{(3)})+\eta_i^{(1,2)}(t).
\end{eqnarray}
The interactions are characterized by coupling parameters
$A^{(1,2)}$ and $B$ to quantify respectively the interactions
within (intra--), and between (inter--), the ensembles; $f$ and
$g$ are $2\pi$-periodic functions that describe coupling in the
ensembles. The fact that $A^{(1)} \neq A^{(2)}$ implies that
the oscillators in the ensembles are asymmetrically coupled.
$\theta_i^{(1,2)}$ are the phases of the $i$th oscillator in
each ensemble and $N^{(1,2)}$ refer to the ensemble sizes; we
take $N^{(1)}=N^{(2)}=N$. From Eq. (\ref{mod01}), it is obvious
that each oscillator will run at its own characteristic
frequency $\omega_i$ when uncoupled. However when coupled,
there tends to arise a collective behavior in the ensemble.
Depending upon the strength of the coupling parameters, the
oscillators either partially or completely synchronize. The
emergence of synchronization is spontaneous beyond a critical
value of the coupling parameter.

The $\eta_i^{(1,2)}$ are independent Gaussian white noises with
$\langle\eta_i^{(1,2)}(t)\rangle =0$ and $\langle\eta_i^{(1,2)}
(t)$ $\eta_j^{(1,2)'}(t) \rangle$
$=2K^{(1,2)}\delta(t-t')\delta_{ij}$ and $K^{(1,2)}$ are the
noise intensities; $K^{(1)} \neq K^{(2)}$ represents noise
asymmetry. Phase asymmetry is introduced by phase shifts
$0\leq\alpha^{(1,2,3)}<\pi/2$. The primary effect of the phase
asymmetry is to synchronize the oscillators to an entrainment
frequency that differs from a simple average of their natural
frequencies. Such asymmetry is widespread in natural systems
like heart cells \cite{Winfree:80} and the cardiorespiratory
interactions \cite{Stefanovska:99,Palus:03}. Phase asymmetry is
used to model synaptic information and time delays in neuronal
networks and also in the phase reduction of nonisochronous
oscillators \cite{Pikovsky:01}. The natural oscillator
frequencies $\omega_i^{(1,2)}$ are assumed to be Lorentzianly
distributed as $g^{(1,2)}(\omega) =
\frac{\gamma}{\pi}(\gamma^2+(\omega^{(1,2)}-\bar{\omega}^{(1,2)})^2)^{-1}$
with central frequencies $\bar\omega^{1,2}$, and $\gamma$ is
the half-width at half-maximum.

\subsection{The Mean Field}
When $N\rightarrow\infty$ in the thermodynamic limit, each
oscillator in the ensemble can be regarded as being coupled to
the mean field. Thus for infinitely many oscillators,
synchronization can conveniently be defined and characterized
by a mean-field (order) parameter
\begin{eqnarray}
r^{(1,2)}e^{i\psi^{(1,2)}}=\frac{1}{N}\sum_{j=1}^N
e^{i\theta_j^{(1,2)}}. \nonumber
\end{eqnarray}
Here $\psi^{(1,2)}(t)$ are the average phases of the
oscillators in the respective ensembles and $r^{(1,2)}(t)$
provide measures of the coherence of each oscillator ensemble,
which varies from 0 to 1. The amplitude of each order
parameters $r^{(1,2)}$ vanishes when the oscillators in the
corresponding ensemble fall out of synchronization with each
other, and is positive for synchronized states, thus
characterizing intra-ensemble synchronization. When
$\delta\psi=\psi^{(1)}-\psi^{(2)}\approx$ constant the
ensembles are mutually locked in phase, defining the state of
inter-ensemble synchronization. Geometrically, if we consider
the phases of all the oscillators to be moving on the unit
circle, then the mean field is the centroid of all the phases.
With this characterization, we show that an increase of the
coupling strength between two ensembles that are synchronized
separately does not immediately result in their mutual
phase-locking. Rather, phase-locking occurs through either one
of two different routes: in Route-I the oscillators in the two
ensembles combine and form clusters; in Route-II one of the
ensembles desynchronizes while the other remains synchronized.
Further, there also exists the possibility that phase-locking
between the ensembles cannot occur at all.

\section{Stability of the incoherent state in the thermodynamic limit}
\label{Stability} In the limit $N\rightarrow \infty$, a density
function can be defined as
$\rho^{(1,2)}(\theta,t,\omega)d\omega d\theta$, to describe the
number of oscillators with natural frequencies within
$[\omega,\omega+d\omega]$ and with phases within
$[\theta,\theta+d\theta]$ at time $t$. For fixed $\omega$ the
distribution $\rho^{(1,2)}(\theta,t,\omega)$ obeys the
evolution equation
\begin{eqnarray}
\frac{\partial\rho^{(1,2)}}{\partial
t}=-\frac{\partial}{\partial\theta}(\rho^{(1,2)}v^{(1,2)})
+K^{(1,2)}\frac{\partial^2\rho^{(1,2)}}{\partial\theta^2}.
\nonumber
\end{eqnarray}
where $v^{(1,2)}$ are given by
\begin{eqnarray}
\small
v^{(1,2)}&=&\omega^{(1,2)} - A^{(1,2)}\int_0^{2\pi}d\theta \int_{-\infty}^{\infty}
 g^{(1,2)}(\omega)\nonumber \\
 & \times &f(\theta-\phi+\alpha^{(1,2)})\rho^{(1,2)}(\phi,t,\omega)d\omega-
B\int_0^{2\pi}d\theta \nonumber \\
 & \times&\int_{-\infty}^{\infty}
 g^{(2,1)}(\omega)h(\theta-\phi+\alpha^{(3)})\rho^{(2,1)}(\phi,t,\omega)d\omega.
\nonumber
\end{eqnarray}
The function $\rho^{(1,2)}(\theta,t,\omega)$ is real and $2\pi$
periodic in $\theta$, so it can be expressed as a Fourier series
in $\theta$
\begin{eqnarray*}
\rho^{(1,2)}(\theta,t,\omega)&=&\sum_{l=-\infty}^{\infty}\rho_l^{(1,2)}(\omega,t)e^{il\theta}\\
&=&\frac{1}{2\pi}+\rho_1^{(1,2)}e^{i\theta}+\mbox{c.c}+\eta(\theta,t,\omega),
\end{eqnarray*}
where c.c is the complex conjugate of the preceding term and
$\eta(\theta,t,\omega)$ denotes the 2nd and higher harmonics.
Substituting $\rho^{(1,2)}(\theta,t,\omega)$ into the evolution
equation, we get
\begin{eqnarray}
\label{any05a}
\dot{\rho_l}^{(1,2)}&+&(il\hat{\omega}^{(1,2)} +l^2K^{(1,2)})\rho_l^{(1,2)}
\nonumber\\
&=&2il\pi\sum_{k=1}^{\infty}(a_k\rho_{l-k}^{(1,2)} +a_k^{\ast}\rho_{l+k}^{(1,2)}),
\end{eqnarray}
where $\rho_{-l}^{(1,2)}=\rho_l^{\ast(1,2)}$, $\hat{\omega}^{(1,2)}=\omega^{(1,2)}-(A^{(1,2)}f_0+Bh_0)$ and $a_k=(A^{(1,2)}e^{ik\alpha^{(1,2)}}f_k\langle
\rho_k^{(1,2)}\rangle +Be^{i\alpha^{(3)}}h_k\langle \rho_k^{(2,1)}\rangle)$. The linearized form of Eq.
(\ref{any05a}) reads as
\begin{eqnarray}
\label{any05}
 \dot{\rho_k}^{(1,2)}&=&-(ik\hat{\omega}^{(1,2)}+k^2K^{(1,2)})\rho_k^{(1,2)}+ika_k,
\end{eqnarray}
\noindent where the Fourier components for $|l|>k$ are
neglected since $l=\pm k$ are the only nontrivial unstable
modes, $\rho_0=1/2\pi$ is the trivial solution corresponding to
incoherence, and $f_k$ and $h_k$ are coefficient of the Fourier
series of functions $f$ and $h$. Here $\langle\cdot\rangle$
represents the average over the frequencies $\omega^{(1,2)}$
weighted by the Lorentzian distribution $g^{(1,2)}(\omega)$.
Solving Eq. (\ref{any05}) we get
\begin{eqnarray}
\label{any07}
\rho_k^{(1,2)}=b_k^{(1,2)}(\omega)e^{\lambda_k t}+O(|\rho|).
\end{eqnarray}
Substituting the above equation back into Eq. (\ref{any05}) we find
\begin{eqnarray}
\label{any05cc}
 b_k^{(1,2)}(\omega)&=&\frac{
(\bar{A}^{(1,2)}\langle b_k^{(1,2)}\rangle +\bar{B}\langle b_k^{(2,1)}\rangle)}{(\lambda_k+ik\hat{\omega}^{(1,2)}+k^2K^{(1,2)})},
\end{eqnarray}
where $\bar{A}^{(1,2)}=ikA^{(1,2)}f_k e^{ik\alpha^{(1,2)}},\;
\bar{B}=ikBh_k e^{ik\alpha^{(3)}}$. The integrals in this
equation can be written as constants $C^{(1,2)}$ which are to
be determined in a self-consistent manner. Thus, for the assumption
\begin{eqnarray}
\label{any05d}
 C_k^{(1,2)}=\int_{-\infty}^{\infty}b_k^{(1,2)}(\omega')g^{(1,2)}(\omega')d\omega'
\end{eqnarray}
Eq. (\ref{any05cc}) for $b_k^{(1,2)}$ becomes
\begin{eqnarray}
\label{any05e}
b_k^{(1,2)}(\omega)=\frac{
(A^{(1,2)}C_k^{(1,2)}+BC_k^{(2,1)})}{(\lambda_k+ik\hat{\omega}^{(1,2)}+k^2K^{(1,2)})}.
\end{eqnarray}
This on substitution back into Eq. (\ref{any05cc}) results in the following 
characteristic equation
\begin{eqnarray}
1&=&\bar{A}^{(1)}m_1
+\bar{A}^{(2)}m_2-(\bar{A}^{(1)}\bar{A}^{(2)}-\bar{B}^2)m_1m_2,
\label{any05b}
\end{eqnarray}
where $m_i=\int_{-\infty}^{\infty}(g^{(i)}(\omega)d\omega)/(\lambda_k
+ik\hat{\omega}^{(i)}+k^2K^{(i)})$, $i=1,2$.
The eigenvalues obtained from (\ref{any05b}) are
\begin{eqnarray}
\label{any05c} \lambda_{k\pm}& =&\frac{\bar{A}^{(1)}+\bar{A}^{(2)}}{2}-\gamma-k^2\bar{K}-ik\bar{\omega}
\pm\frac{1}{2}\bigg(4\bar{B}^2+\Delta A^2\nonumber\\
&-&k(\Delta \omega-ik\Delta K)
(\Delta \omega-ik\Delta K+2i\Delta A)
\bigg)^{\frac{1}{2}},
\end{eqnarray}
where $\bar{K}=\frac{(K^{(1)}+K^{(2)})}{2}$, $\Delta K=K^{(1)}-K^{(2)}$, $\Delta\omega=\bar{\omega}^{(1)}-\bar{\omega}^{(2)}$, $\Delta A=(\bar{A}^{(1)}-\bar{A}^{(2)})$
$\bar{\omega}=(\bar{\omega}^{(1)}+\bar{\omega}^{(2)})/2$.

\begin{figure}
\begin{center}
\includegraphics[width=6.5cm]{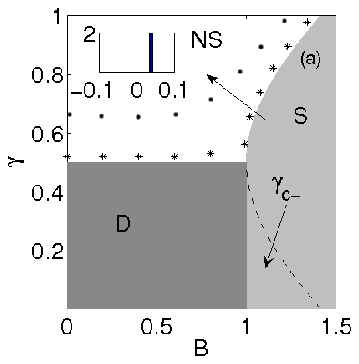}
\includegraphics[width=6.5cm]{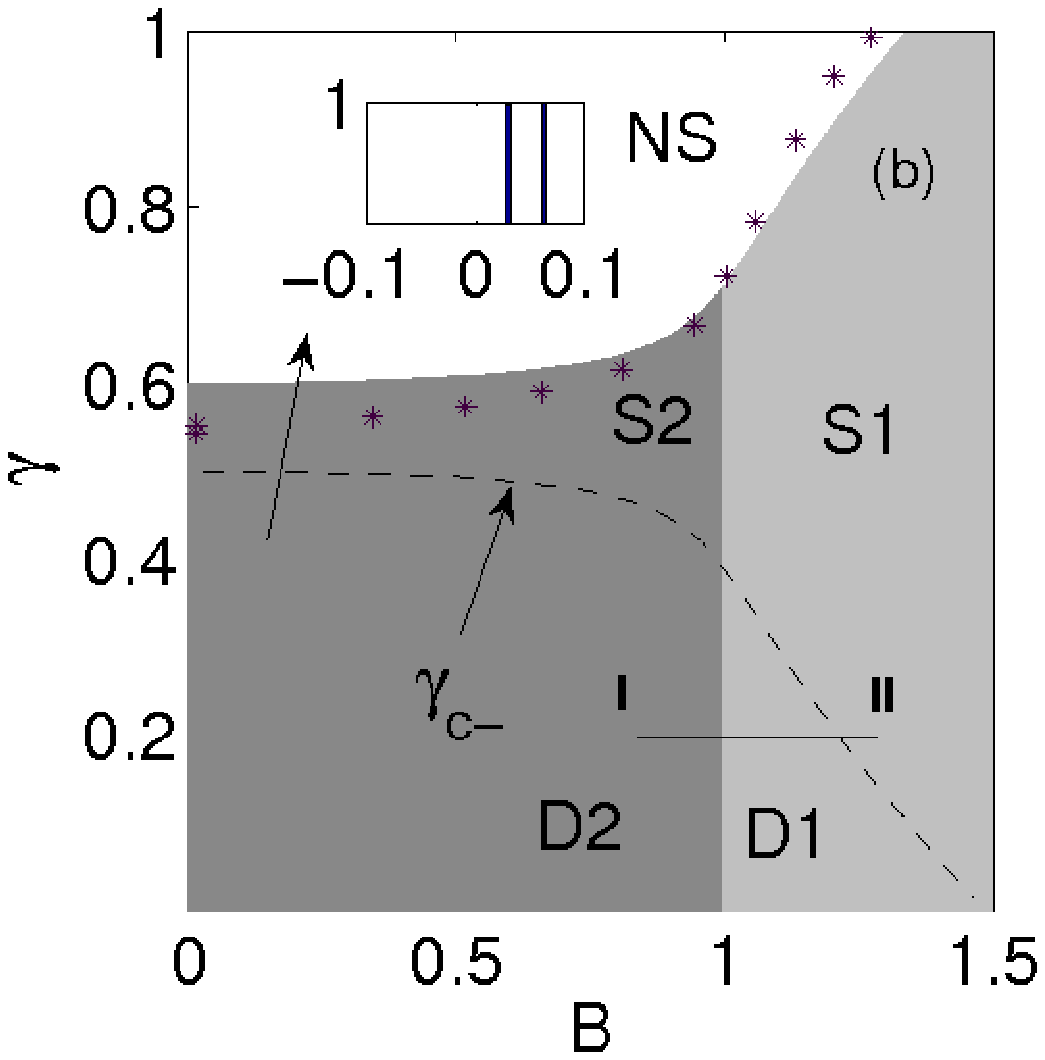}
\caption{Theoretical $B$--$\gamma$ bifurcation diagram for
$\alpha=0,\Delta\omega=1,K^{(1,2)}=0$, with (a)
$A^{(1)}=A^{(2)}=1$, and (b) $A^{(1)}=1.2$, $A^{(2)}=1.0$. The
different synchronization regimes are as follows: NS, no
synchronization; S, synchronization with single entrainment
frequency (reached via a single Hopf bifurcation); D,
synchronization with two entrainment frequencies (two Hopf
bifurcations); S1, both the ensembles entrained to a single
frequency; S2, either of one of the ensembles synchronized with
single entrainment frequency; D1, the two ensembles behave as
one, with the oscillators in each ensemble entrained to either
of the two distinct frequencies; D2, synchronization in both
the ensembles separately with two entrainment frequencies.
Regardless of symmetry, the notations S and D represent
respectively single or double frequencies occurring after one
or two Hopf bifurcations. The boundary between regimes NS and
S/D represents $\gamma_{c+}$. The lines of $*$s represent the
numerically determined bifurcation boundaries for $r>0.7$. For 
comparison, a numerical boundary ($\bullet$s) for $r>0.5$ is also plotted in (a). 
Insets show the
frequency distributions (also obtained numerically) for the
indicated regions; their ordinate axes represent oscillator
counts in thousands. Note that the occurrence of perfect
synchronization with (a) $2000$ and (b) $1000$ oscillator
groups cannot be expected throughout the whole of each
indicated region. The line I--II in (b) is one of the routes 
to synchronization discussed in Sec. \ref{routes}.} \label{gmp1}
\end{center}
\end{figure}

For a detailed analysis of the above equation, we specify
sinusoidal forms for the functions $f$ and $h$ as
$\{f,h\}(\theta)=\sin\theta$. Therefore the eigenvalue equation
(\ref{any05c}) becomes
\begin{eqnarray}
\lambda_{\pm}&=&
-\gamma-\bar{K}+\frac{\kappa}{4}e^{i\alpha}\pm\frac{1}{2}(\xi e^{2i\alpha}
-\hat{A}(\Delta K+i\Delta\omega) e^{i\alpha}
\nonumber\\&&-\Delta\omega^2+\Delta K^2+2i\Delta\omega\Delta K)^{\frac{1}{2}}
-i\bar{\omega},
 \label{tany06}
\end{eqnarray}
or equivalently we have
\begin{eqnarray}
\lambda_{\pm} = \left\{
\begin{array}{ll}
-\bar{K}-\gamma+\frac{\kappa}{4}e^{i\alpha}\pm\frac{1}{2}(p^2+q^2)^{\frac{1}{4}}
e^{i\frac{1}{2}\zeta}-i\bar{\omega},&\\
\qquad\qquad\qquad\qquad\qquad\qquad\qquad p>0&
\\\\
-\bar{K}-\gamma+\frac{\kappa}{4}e^{i\alpha}\pm\frac{i}{2}(p^2+q^2)^{\frac{1}{4}}
e^{i\frac{1}{2}\zeta}-i\bar{\omega},&\\
\qquad\qquad\qquad\qquad\qquad\qquad\qquad p<0&
\end{array}
\right. \label{any06}
\end{eqnarray}
where $\alpha_i=\alpha$, $i=1,2,3$, $\kappa=A^{(1)}+A^{(2)}$, $\hat{A}=(A^{(1)}-A^{(2)})$,
$\xi=(\frac{1}{4}\hat{A}^2+B^2)$,
$\zeta=\tan^{-1}(\frac{q}{p})$,
$p=\xi\cos(2\alpha)+\hat{A}[\Delta\omega\sin{\alpha}-\Delta K
\cos\alpha]-\Delta\omega^2+\Delta K^2$,
$q=\xi\sin(2\alpha)-\hat{A}\Delta\omega\cos{\alpha}-\hat{A}\Delta
K\sin{\alpha}+2\Delta\omega\Delta K$.
The resultant bifurcation diagram is shown in Fig. \ref{gmp1}.
It is discussed in detail below, in Secs. \ref{Coupling} and
\ref{Phase}. Here it is obvious that for the case when phase asymmetry is absent, when
$A^{(2)}=B=0$, the characteristic
equation (\ref{any05b}) reduces to the characteristic
equation of the Kuramoto model derived by
Strogatz et. al. \cite{Kuramoto:84,Strogatz:00}.

\subsection{Numerical considerations}
\label{NumMtds}

To investigate the system numerically, we use a Runge-Kutta
fourth order (RK4) routine for solving the model equations with
a time step of 0.01 (we have confirmed that the results are not
affected by decreasing the time step below 0.01). We take $N=1000$ 
in each ensemble and the initial phases of the oscillators 
are assumed to be equally distributed within the interval 
$[0,2\pi]$. As a signature of
synchronization, we take the condition $Re(\lambda_{\pm})>0$ in
the case of the analytic treatment. For the numerical
experiment, we set $r^{(1,2)}>0.7$ for intra-ensemble
synchronization in the corresponding ensembles, and a constant
$\delta\psi$ for inter-ensemble synchronization as the
conditions. The numerical condition for intra-ensemble
synchronization, that $r^{(1,2)}>0.7$, may at first seem too
strict when compared with the analytic condition that
$r^{(1,2)}>0$. However, there are certain differences between
analytic and numeric considerations that make this choice
reasonable. Mainly, $N$ is finite for the numerical experiment,
whereas analytic conditions are derived in the limit
$N\rightarrow\infty$. Further, the analytic and numeric
bifurcation boundaries (discussed later) are found to match
quite closely for this choice of the numeric threshold for
$r^{(1,2)}$. We have plotted the numerical
boundary for $r=0.5$ along with $r=0.7$ in Fig. \ref{gmp1}(a)
to illustrate this.

\section{Synchronization regimes}
\label{SychRegimes}

We have identified analytically the possibility of five
distinct dynamical regimes \cite{Jane:08a}:
\begin{itemize}

\item NS: the region of no synchronization or incoherence
    (steady state).

\item S1: the region of global (inter-ensemble)
    synchronization, in which the oscillators of both
    ensembles are all entrained to the same frequency.

\item S2: the region where there is synchronization within
    one ensemble but not the other.

\item D2: the region of synchronization within both
    ensembles, separately and independently, with two
    different entrainment frequencies.

\item D1: a global regime in which the two ensembles behave
    as one, but the oscillators within each ensemble are
    entrained at either one of two distinct entrainment
    frequencies. We will call this phenomenon {\it
    inter-ensemble clustering}.

\end{itemize}

\noindent Regions S2 and D1 cannot occur when coupling and
noise asymmetries are absent \cite{Okuda:91,Montbrio:04} (see
Fig. \ref{gmp1} (a)). In the following subsections, we will
consider how these synchronization regimes are affected by
coupling, noise and phase asymmetries, respectively.

\subsection{The effect of coupling asymmetry}
\label{Coupling} Consider Fig. \ref{gmp1}(b) for the case
$\alpha=0$ and $K^{(1)}=K^{(2)}=0$, when $\xi-\Delta\omega^2>0$. If we
start from the state of no synchronization (region NS), and
decrease $\gamma$ for fixed $B>1$, the
incoherent (steady) state becomes unstable via a single Hopf
bifurcation. Thus the system enters into the region S1 from NS
(crossing $\gamma_{c+}$) and the ensembles entrain to a single
frequency $\Omega_+$. With further decrease of $\gamma$ below
the $\gamma_{c-}$ line in the region D1 (in Fig. \ref{gmp1}), a
new entrainment frequency emerges through a second Hopf
bifurcation. In this region, the oscillators from the two
ensembles combine and form two clusters (inter-ensemble
clustering) oscillating with two frequencies,
\begin{eqnarray}
\Omega_{\pm}&=&-\mbox{Im}(\lambda_{\pm})=\pm (1/2)\left[(\xi-\Delta\omega^2)^2
 +\hat{A}^2\Delta\omega^2\right]^{\frac{1}{4}} \nonumber \\
 && \quad \times \sin{\left(\frac{1}{2}\tan^{-1}
 \left[\hat{A}\Delta\omega/(\xi-\Delta\omega^2)\right]\right)}+\bar{\omega}.
 \label{D1freq01}
\end{eqnarray}
The lines $\gamma_{c\pm}$ in Fig. \ref{gmp1} are obtained by imposing
the condition $Re(\lambda_{\pm})=0$ in Eq. (\ref{any06}).

\begin{figure}
\begin{center}
\includegraphics[height=6cm,width=7.5cm]{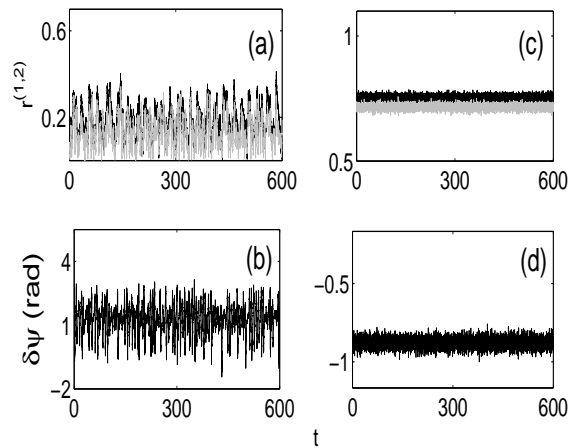}
\caption{Time variations of the coherence parameters $r^{(1)}$
(grey), $r^{(2)}$ (black) and the phase difference
$\delta\psi$, as obtained from numerical simulations. Parameter
values are: (a), (b) $B=1$, $\alpha=0.23$ and (c), (d) $B=1$,
$\alpha=0.47$ corresponding to regions D1 and D2 respectively
of Fig. \ref{gmp1}(b) as traveling along the line $I-II$. Note
that in region D1 the order parameters display no
synchronization.}\label{gmp5}
\end{center}
\end{figure}

\begin{figure}
\begin{center}
\includegraphics[height=3.5cm,width=7.5cm]{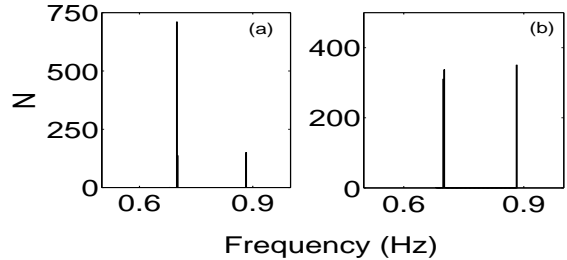}

\caption{The distribution of frequencies in region D1 for the
same parameter values as in Fig. \ref{gmp5}(a): (a) first
ensemble; (b) second ensemble. The splitting of the first
frequency component into two almost indistinguishable
sub-components corresponds to a discrepancy between numerics
and analytics, attributable to approximations (see text) made
in the former.}\label{gmp7}
\end{center}
\end{figure}

Thus in this region the order parameters $r^{(1,2)}$ either
fluctuate in a quasi-periodic manner or have complicated
dynamics (see Figs.\ \ref{gmp5}(a) and (c)). This is because
each ensemble has two clusters oscillating with different
frequencies (see Figs.\ \ref{gmp5}(a) and \ref{gmp7}). Thus,
the behavior of the order parameters $r^{(1,2)}$ is quite
subtle. Since $r^{(1,2)}$ measure only the amount of
synchronization within an ensemble, a decrease in $r^{(1,2)}$
will not necessarily mean desynchronization. Rather, for a
sufficient value of the coupling parameters, a decrease in
$r^{(1,2)}$ represents  a signature of inter-ensemble
synchronization: clustering corresponds to the occurrence of
desynchronization within an ensemble because some of its
oscillators tend to synchronize with the other ensemble.

Again looking at Fig. \ref{gmp1}(b), when
$B<1$, a decrease in $\gamma$ takes the system from region NS
to region S2 by crossing the line $\gamma_{c+}$ through a
single Hopf bifurcation. Further decrease in $\gamma$ causes
the system to cross the line $\gamma_{c-}$ and, via another
Hopf bifurcation, enter into region D2 where there are two
entrainment frequencies $\Omega_{\pm}$. The latter can be
calculated from Eq.\ (\ref{any06}). In region S2,
intra-ensemble synchronization can occur in either one of the
ensembles, depending upon whether $A^{(1)}$ or $A^{(2)}$ is
greater; in Fig. \ref{gmp1}(b), since $A^{(1)}>A^{(2)}$,
synchronization occurs in the first ensemble with the second
ensemble remaining incoherent. Note that, on increasing $B$
(for fixed $\gamma$) while in region S2, the condition
$\xi-\Delta\omega^2<0$ is violated and the ensembles enter into
the phase-locked region S1. In region D2, the ensembles
synchronize separately to two the locking frequencies (unlike region D1 where the ensembles
combine and synchronize to two locking frequencies given by Eq.
(\ref{D1freq01})).

The corresponding $(B-\gamma)$ bifurcation diagram for the case
$A^{(1)}$=$A^{(2)}$ is plotted in Fig. \ref{gmp1}(a) to show
the difference between these two cases. Region D represents
intra-ensemble synchronization which occurs through a
degenerate Hopf bifurcation (similar to the route to D2) with
entrainment frequencies
$\Omega_{\pm}=\mp(1/2)(\Delta\omega^2-B^2)^{\frac{1}{2}}
+\bar{\omega}$ and S represents inter-ensemble synchronization
through a single Hopf bifurcation (similar to the route to S1)
with same frequency $\Omega=\bar{\omega}$. Note that regions S2
and D1 cannot arise for the symmetric coupling case and that
these two synchronization regimes are therefore induced by
coupling asymmetry.

The presence of two entrainment frequencies in region D1 can be
seen by looking at the frequencies into which all the
individual oscillators are grouped as shown in Fig. \ref{gmp7}
(since the order parameters do not reveal this synchronization
phenomenon). The inter-ensemble clustering that occurs in this
case is quite different from the formation of clusters in a
single ensemble \cite{Kuramoto:84,Tass:97,Kouomou:03} -- here
the oscillators in two different ensembles \emph{combine} and
form clusters. The occurrence of this phenomenon provides a new
insight into possible ways of controlling synchronization in
more realistic situations (considering asymmetry) like neural
networks where some neurons from one ensemble (say cortex) tend
to synchronize with other ensemble (say thalamus) creating
desirable (temporal coding) or undesirable effects (as in the
case of epileptic seizures). For instance, in a thalamocortical
model of the neuronal synchronization mechanisms during
an{\ae}sthesia \cite{Jane:08b}, we found that the transition
from deep to light an{\ae}sthetized state occurs as a result of
a fraction of the thalamic neurons entering into
synchronization with the cortex, at the same time losing
synchronization within its own ensemble. The clustering that
occurs in this case is desirable in the sense that it favours
coding of sensory information and helps the brain to resist the
effects of an{\ae}sthesia and successfully maintain
consciousness and cognition. Without coupling asymmetry, these
phenomena would not occur.

\subsection{The effect of noise asymmetry}
\label{Noise}

It is well known that real physical systems are in general
subject to noise. Here, we regard as ``noise'' any kind of
random fluctuation in the system, whether originating
internally or externally.
Synchronization effects, induced and modified noise, are one of
particular interest
\cite{He:03,Goldobin:05,Nakao:07,Kawamura:07,Kawamura:08}.

\begin{figure}
\begin{center}
\includegraphics[width=7.5cm]{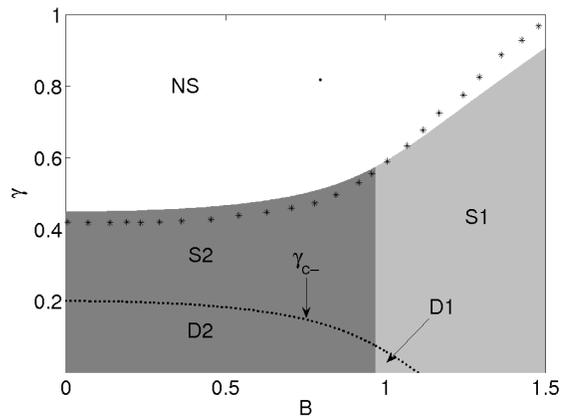}
\caption{Theoretical $B$--$\gamma$ bifurcation diagram for
asymmetric noise with $A^{(1)}=A^{(2)}=1$, $\Delta K=0.25$ with 
$K^{(1)}=0.3$ and $K^{(2)}=0.05$,
$\Delta \omega=1$. Note that the synchronization regimes S2 and
D1 emerge in the presence of noise asymmetry even for symmetric
coupling (cf.\ Fig. \ref{gmp1}(a)).} \label{N_bifanal1}
\end{center}
\end{figure}

\begin{figure}
\begin{center}
\includegraphics[width=8.5cm]{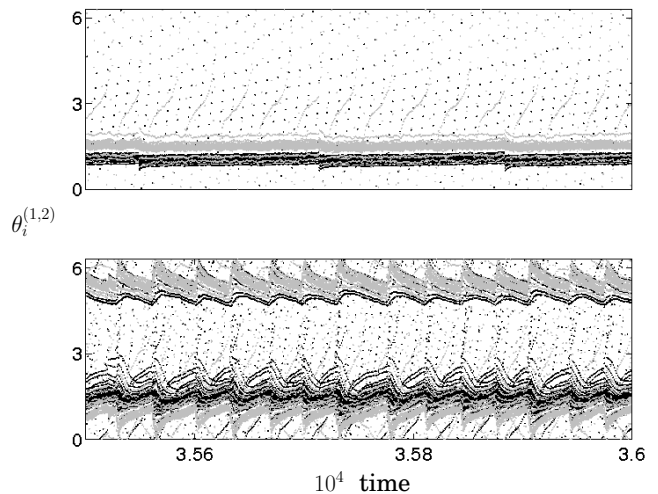}
\caption{Time evolution of the oscillator phases in the first
(grey) and second (black) ensembles. Parameter values are
$A_1=A_2=1.4$, $B=1$, $\gamma=0.05$, and either $\Delta K=0$,
with $K^{(1)}=0.2$ and $K^{(2)}=0.2$ (top) or $\Delta K=0.25$,
with $K^{(1)}=0.2$ and $K^{(2)}=-0.05$  (bottom). 
Thus the top and the bottom
panels represent respectively the synchronization regions S and
D1, induced by noise asymmetry.} \label{N_Num1}
\end{center}
\end{figure}

When asymmetric noise in introduced into a system with
asymmetric coupling, the bifurcation regimes remain the same in
the presence of coupling and phase asymmetries. There may be
changes in the boundaries of the respective regions and their
entrainment frequencies. However, for the case of symmetric
coupling, asymmetric noise can induce the phenomenon of global clustering. We
have already seen in Sec. \ref{Coupling} that inter-ensemble
clustering phenomena (region D1) cannot occur in a system with
symmetric coupling and symmetric noise.
Fig. \ref{N_Num1} plots the individual oscillator phases,
determined numerically, indicating the transition from S1 to D1
induced by noise asymmetry. When $A^{(1)}=A^{(2)}=1.4$, $B=1$,
$\gamma=0.05$ and $\Delta K=0$ the system is in region S
(corresponding to Fig. \ref{gmp1} (a)) where synchronization occurs
in both the ensembles with one entrainment frequency. This can
be seen from the top panel of Fig.\ \ref{N_Num1} where
oscillators from both ensembles lock to form a single major
cluster. On the other hand, when $\Delta K=0.25$, the combined
system of the two ensembles synchronize to two main clusters,
each of which comprises a fraction of the oscillators from both
ensembles (see Fig. \ref{N_Num1} (bottom)), representing region
D1. Thus it is becomes obvious that asymmetric noise can in
some ways imitate the effects of asymmetric coupling when the
latter is absent.

\begin{figure}
\begin{center}
\includegraphics[width=7.5cm]{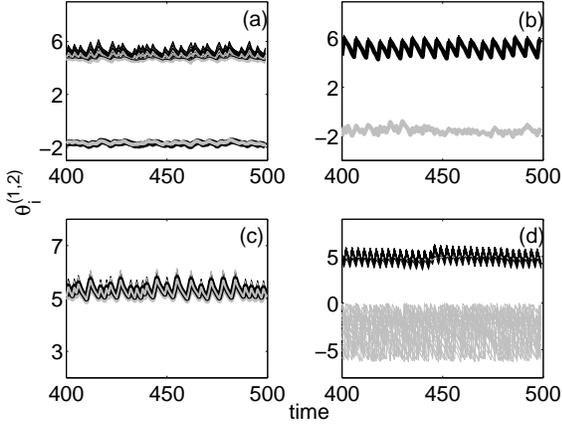}
\caption{Noise asymmetry-induced synchronization regimes
obtained numerically for the case of symmetric coupling. Black
and grey lines represent the time evolution of the oscillator
phases in the first and second ensembles respectively, for the
same parameter values as in Fig.\ \ref{N_bifanal1} and: (a)
$B=1$, $\gamma=0.1$; (b) $B=0.5$, $\gamma=0.1$; (c) $B=1.2$,
$\gamma=0.4$; and (d) $B=0.6$, $\gamma=0.4$. Panels (a)--(d)
represent the synchronization regimes D1, D2, S1, S2
respectively.} \label{Balpha_num5}
\end{center}
\end{figure}

The S2 region also appears in this case, induced by noise
asymmetry. Here too, depending upon whether $\Delta K$ is
positive or negative, synchronization occurs either in the
second or the first ensemble respectively, similar to the case
when region S2 arises in the presence of coupling asymmetry.
Thus noise asymmetry plays a similar role to coupling asymmetry
for the symmetric coupling case, and the $(B-\gamma)$
bifurcation diagrams \ref{gmp1}(b) and \ref{N_bifanal1} look
similar. Fig. \ref{Balpha_num5} depicts the results of
numerical investigation of all the synchronization regimes in
the presence of noise asymmetry corresponding to the analytical
bifurcation diagram in Fig. \ref{N_bifanal1}. In contrast, for
symmetric noise the dynamics is unaffected, no matter whether
coupling and phase asymmetries are present or absent. The only
difference is that the incoherent state becomes unstable for
larger values of the critical parameters as one increases noise
intensity.

\subsection{The effect of phase asymmetry}
\label{Phase} 
\begin{figure}
\begin{center}
\includegraphics[width=7.5cm]{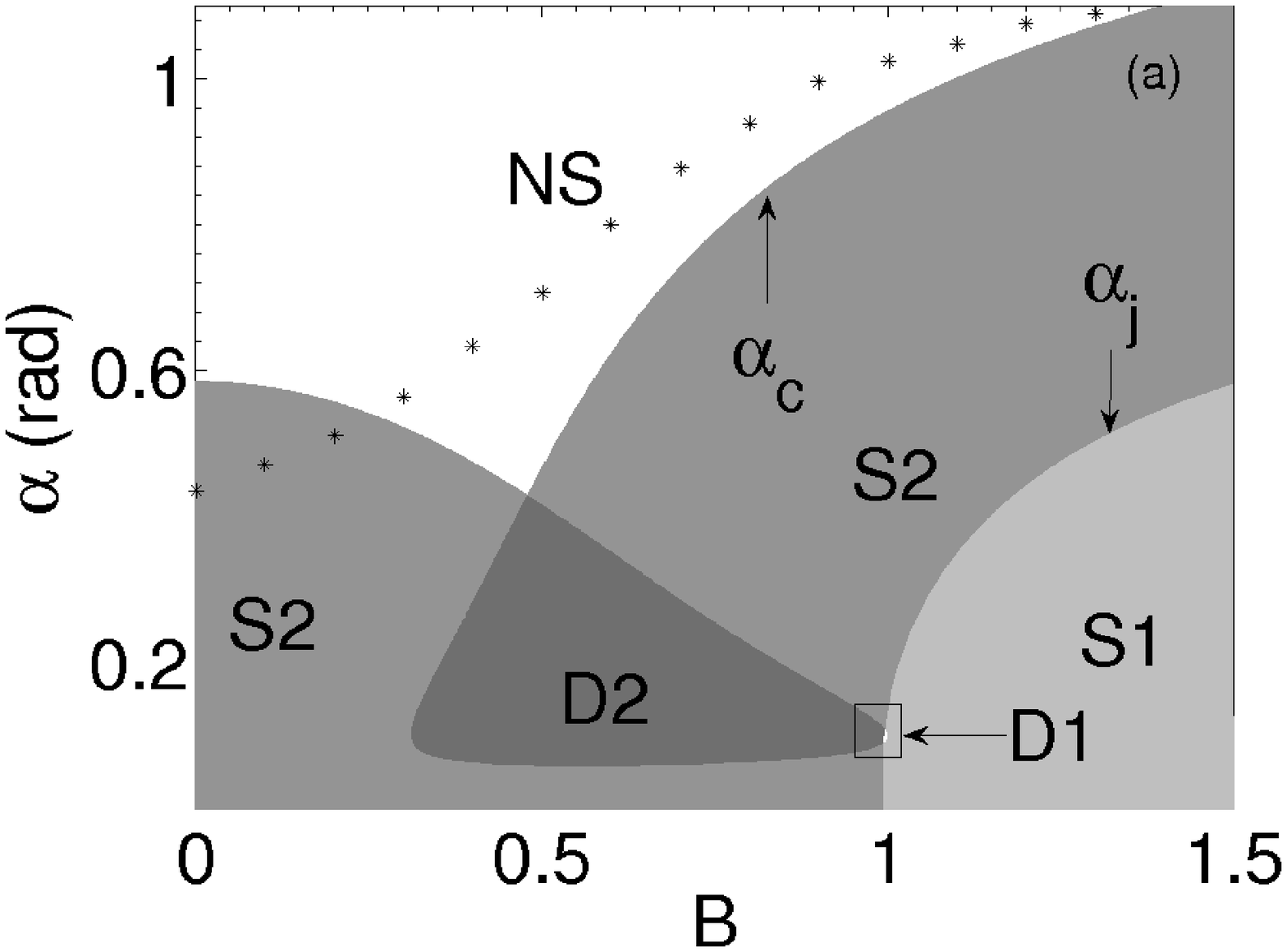}
\includegraphics[width=7.5cm]{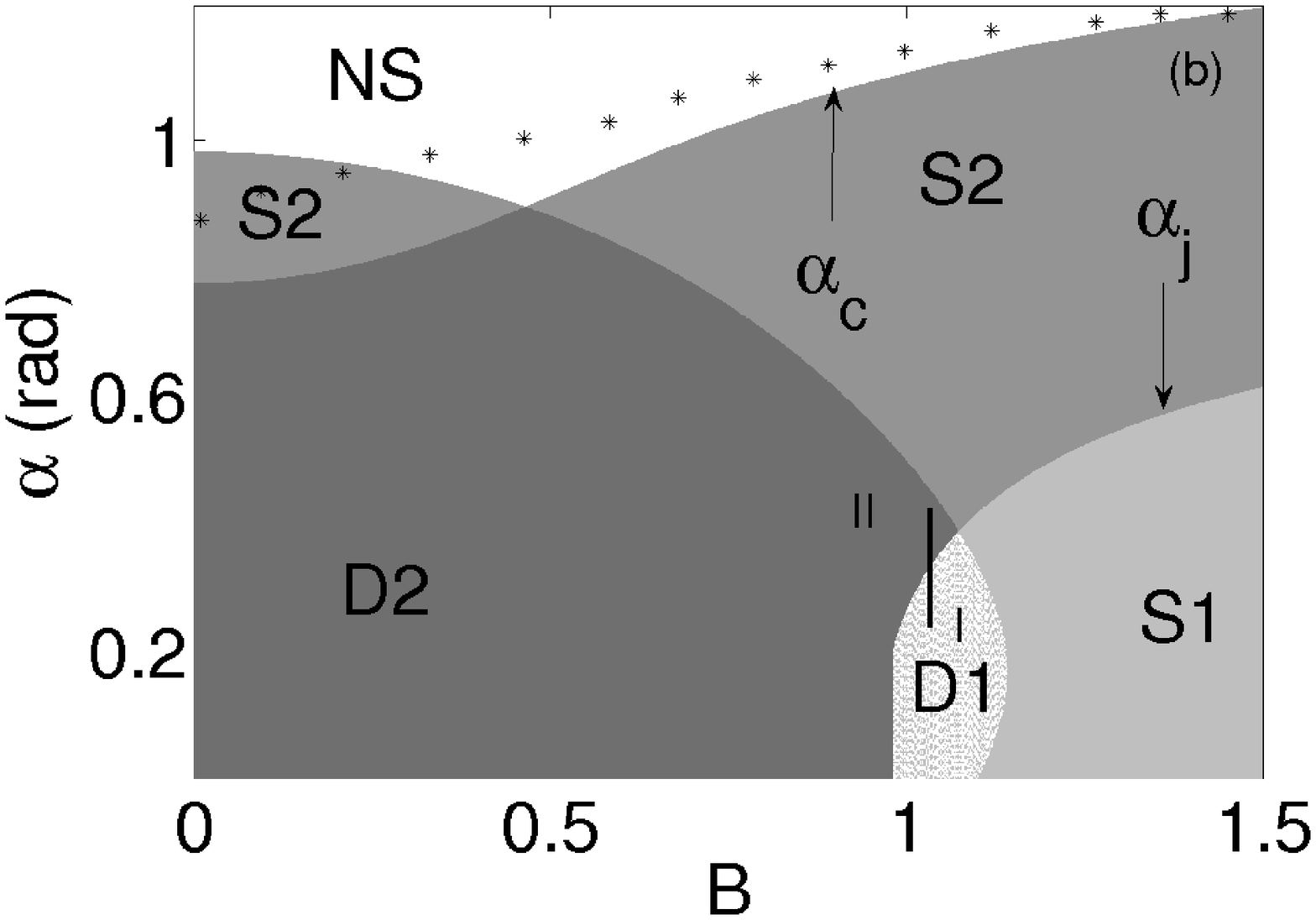}
\caption{Theoretical $B$--$\alpha$ bifurcation diagram for (a)
$A^{(1)}=1.2,A^{(2)}=1$, (b) $A^{(1)}=1.8,A^{(2)}=1.4$ and
$\Delta\omega=1,\gamma=0.5$. The line of $*$s represents the
numerically obtained bifurcation boundary between the
synchronized and incoherent states. Greatly reduced S1 and D1
regions occur due to the presence of phase asymmetry. The
discrepancy between numerical and analytic boundaries is
discussed in the text.} \label{gmp2}
\end{center}
\end{figure}

For the case $\alpha\ne0$, the inter-ensemble
regions D1 and S1 shrink as $\alpha$ increases, whereas the
intra-ensemble synchronization region S2 expands, as shown in
Fig. \ref{gmp2}. This means that finite phase asymmetry reduces
the probability of inter-ensemble synchronization (note reduced
S1 and D1 regions in Fig. \ref{gmp2}) and mostly allows only
intra-ensemble synchronization of one or both of the ensembles.
For a given set of parameters, on increasing $\alpha$ from 0,
the following condition is satisfied
\begin{eqnarray}
\xi\cos(2\alpha)
+\hat{A}\Delta\omega\sin{\alpha}-\Delta\omega^2>0
\label{Pcont01}
\end{eqnarray}
up to a critical value of $\alpha=\alpha_j$ given by
\begin{eqnarray*}
\alpha_j=\sin^{-1}\bigg[\frac{\hat{A}\Delta\omega\pm(8\xi^2
-\Delta\omega^2(\hat{A}^2+B^2))^{\frac{1}{2}}}{4\xi}\bigg],
\end{eqnarray*}
where again, for a given set of parameters, there can only be
one value of $\alpha$ that satisfies $0<\alpha\leq\pi/2$. Upon
crossing $\alpha_j$, the condition (\ref{Pcont01}) is violated
and the following condition is satisfied
$\xi\cos(2\alpha)+\hat{A}\Delta\omega\sin{\alpha}
-\Delta\omega^2<0$.

As a result, when $\alpha>\alpha_j$ the inter-ensemble
synchronization breaks down and the system enters into a state
of intra-ensemble synchronization. Thus as one travels from S1
(D1) to S2 (D2) across $\alpha_j$ the combined synchronization
with single (double) frequency breaks between the ensembles and
independent synchronization with single (double) frequency
regime appears. Region S2, unlike region S in Fig. \ref{gmp1}
(a), embraces two states (i) synchronization in ensemble 1 with
ensemble 2 incoherent and (ii) synchronization in ensemble 2
with ensemble 1 incoherent, but does not distinguish between
them.

\begin{figure}
\begin{center}
\includegraphics[height=6.5cm,width=7.5cm]{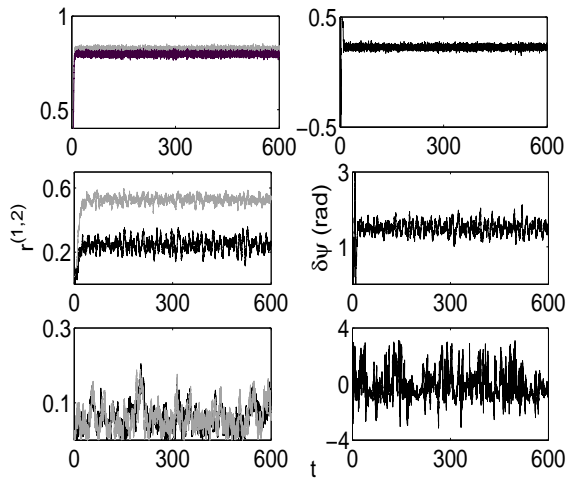}

\caption{The coherence parameters $r^{(1)}$ (grey) and
$r^{(2)}$ (black), and the phase difference $\delta\psi$,
plotted as functions of time, obtained from numerical
simulations. Parameter values were: top panel, $B=0.7,
\alpha=0.2$; middle panel, $B=0.7, \alpha=\pi/4$ (near
$\alpha_c$); and bottom panel, $B=0.7, \alpha=1.2$
($\alpha>\alpha_c$), corresponding to regions D2 near the NS/S2
boundary and NS respectively (see Fig. \ref{gmp2}(a)).
}\label{gmp3}
\end{center}
\end{figure}
Further, there is a critical value of $\alpha=\alpha_c$ above
which the collective oscillations disappear and the incoherent
state becomes stabilized (see Figs. \ref{gmp2} and \ref{gmp3}).
Thus by reducing the chances of occurrence of inter-ensemble
synchronization and favoring intra-ensemble synchronization,
phase asymmetry plays a crucial role in determining the route
to synchronization. Thus, for instance, in a particular
problem, if one wants to have (avoid) the phenomenon of
inter-ensemble clustering (region D1) then it is obvious that
phase asymmetry should be absent (finite, large).

The discrepancy between the numerically and analytically
obtained boundaries in Fig.\ \ref{gmp2} is attributable to the
influence of phase asymmetry. This affects region S2 which is
large here (cf.\ Fig.\ \ref{gmp1}(b) where both S2 and the
discrepancy are smaller) and it changes the thresholds for
$r^{(1)}$ and $r^{(2)}$. Note that neither numerics nor
analytics provides an exact result. The analytic boundary is
obtained from the condition $r^{(1,2)} > 0$ and refers to the
limit of infinitely many oscillators. For numerics, the
(asterisked) boundary is obtained from the condition $r^{(1,2)}
> 0.7$ and refers to a finite number of oscillators.

\subsection{Stability of the fully synchronized states in the limit
$\gamma=0$} In this subsection we focus on the noise-free case
with a frequency distribution that has an infinitely sharp
peak. In this case, the dynamics is reduced to that of two
ensembles of identical oscillators (the oscillators within the
ensembles are identical while the ensembles themselves are
non-identical). Now the $\delta\psi$ corresponding to
intra-ensemble synchronization can be obtained from Eq.\
(\ref{mod01}) as
\begin{eqnarray*}
\delta\psi=\sin^{-1}[(\Delta\omega-\hat{A}\sin\alpha)/2B\cos\alpha].
\end{eqnarray*}
A linear stability analysis of Eq. (\ref{mod01}) then gives $N-1$
degenerate eigenvalues for each ensemble, namely
\begin{eqnarray*}
\lambda_{\pm}=-A^{(1,2)}\cos\alpha-B\cos(\pm\delta\psi+\alpha)<0,
\end{eqnarray*}
that characterize the stability of the intra-ensemble
synchronized states of ensembles 1 and 2 respectively. In
addition, two eigenvalues $\lambda_0=0$ and $\lambda_c=-2B\cos
\alpha \cos \delta\psi$ characterize the stability of
inter-ensemble synchronization. Hence the transition between
inter-ensemble and intra-ensemble synchronization states occurs
at the following bifurcation point
\begin{eqnarray*}
B_c=(\Delta\omega-\hat{A}\sin\alpha)/2\cos\alpha.
\end{eqnarray*}
Note that $\delta\psi$ varies from $-\pi/2$ to $\pi/2$ as
$\alpha$ increases. For $\alpha<\alpha_h$ (not shown in
figures), the stability condition is satisfied by both
ensembles and so intra-ensemble synchronization occurs in both
ensembles. When $\alpha\geq\alpha_h$ the stability condition is
violated by either one of the ensembles and at that point a
Hopf bifurcation occurs. As a consequence, intra-ensemble
synchronization occurs in one of the ensembles. For the case
$A^{(1)}=A^{(2)}$, when $\delta\psi$ varies from $\pi/2$ to $0$
for increasing $\alpha$ and when $\alpha\geq\alpha_h$ the
stability condition is violated by the first ensemble. On the
other hand, when $\delta\psi$ varies from $0$ to $-\pi/2$ with
increasing $\alpha$, the stability condition is violated by the
second ensemble above $\alpha_h$.

\section{Routes to synchronization}
\label{routes}

Given that the system possesses distinct synchronization
regimes, it is of interest to investigate the routes it follows
to synchronization. As one would expect, the route depends on
the coupling, noise and phase asymmetries. In particular, in
the presence of coupling asymmetry, we have identified the
following routes \cite{Jane:08a}, grouped into the two
different cases $\alpha=0$ and $\alpha\neq0$, and assuming that
we increase the inter-ensemble coupling parameter $B$ keeping
all the other parameters fixed. When $\alpha=0$ we find that
there are at least three typical routes:
\begin{enumerate}
\item The oscillators in the ensembles pass from the
    synchronization regime D2 through D1 to the region S1.
    Thus when the ensembles are synchronized separately,
    increasing $B$ results in inter-ensemble clustering
    which then leads to inter-ensemble synchronization or
    phase locking between the ensembles. This route is
    represented by line I-II of Fig. \ref{gmp1}(b).
\item There is also a possibility that when the ensembles
    are synchronized separately and when $B$ is increased,
    the intra-ensemble synchronization be destroyed in one
    of the ensembles, which on further increase of $B$,
    leads to phase-locking between the ensembles. Thus when
    the system is in region D2 increasing $B$ causes the
    system to pass through the region S2 to region S1.
    Inter-ensemble clustering does not occur in this route.
\item If the ensembles are initially not synchronized (that
    is in region NS), then increasing $B$ can cause
    phase-locking of the ensembles directly. Thus the
    system can pass directly from region NS to to S1. This
    route is characteristic of the case $\alpha=0$ and
    cannot occur in the presence of phase asymmetry.
\end{enumerate}
In the presence of phase asymmetry, i.e.\ $\alpha\neq0$, the
ensembles can follow any of the following routes to
synchronization
\begin{enumerate}

\item The ensembles pass from region D2 through D1 to S1.
    This route is similar to route 1 that occurs for the
    case $\alpha=0$. Note that when $A^{(1)}=A^{(2)}$ or
    $\Delta\omega=0$ only one entrainment frequency exists
    below $\gamma_{c-}$ and therefore this route does not
    occur for either cases (due to the non-occurrence of
    region D1).

\item The ensembles pass from region D2 through S2 to S1.
    This route does not incorporate the state of
    inter-ensemble clustering.

\item When the ensembles are synchronized separately (in
    region D2), increasing $B$ causes the disruption of
    synchronization in one of the ensembles leading to
    synchronization in the other ensemble (region S2). Thus
    the ensembles pass from region D2 to S2. This route is
    characteristic of phase asymmetry and cannot occur for
    the case $\alpha=0$.

\item If the ensembles are not synchronized, increasing $B$
    will result in synchronization in either one of the
    ensembles. Thus the ensembles pass from regions NS to
    S2 (unlike NS to S1 in the absence of phase asymmetry).

\end{enumerate}

Knowledge of these routes to synchronization is obviously
important for the control of synchronization in real systems.

\section{Discussion and conclusions}
\label{summary}

One might intuitively suggest that the synchronization
phenomena induced by coupling and noise asymmetries could also
be obtained by choosing a sufficiently large difference between
the mean frequencies of the two ensembles. However, the
synchronization phenomena corresponding to the D1 and S2
regions can only be explained by introducing either coupling or
noise asymmetries. As an illustration let us consider the
eigenvalue for the noise-free case without coupling and phase
asymmetries for $\Delta\omega^2>B^2$
 \begin{eqnarray}
\lambda_{\pm} =
-\gamma+\frac{A}{2}\pm i\frac{1}{2}\sqrt{\Delta\omega^2-B^2}-i\bar{\omega}. \nonumber
\label{illus01}
\end{eqnarray}
For this case, the intra-ensemble synchronization takes place
simultaneously in the two ensembles since the curves
$\gamma_{c+}$ and $\gamma_{c-}$ coincide when
$Re(\lambda_{\pm})$ becomes positive. Although there occur two
Hopf bifurcations, they happen to be one and the same and hence
one will not be able to explain the synchronization region S2.
A similar problem occurs also with the D1 synchronization
regime for $\Delta\omega^2<B^2$. Therefore we must conclude
that the introduction of coupling/noise asymmetries are crucial
to account for certain synchronization phenomena and can never
be replaced by the introduction of large difference between the
mean frequencies of the two ensembles.

It is therefore essential to take account of possible asymmetry
while attempting to model natural systems. Certain phenomena,
like those discussed here, are attributable to asymmetries in
the interactions.

In this paper, we have investigated the role played by
coupling, noise and phase asymmetries in two coupled phase
oscillator ensembles. We have identified a global clustering
phenomenon that may be characteristic of either the coupling or
the noise asymmetry when the other is absent. Phase asymmetry
reduces the likelihood of global clustering and also introduces
new routes that are characteristic of itself. Thus phase
asymmetry controls the routes to inter-ensemble
synchronization. The phenomenon of inter-ensemble clustering
that is characteristic of coupling asymmetry is found to occur
even for symmetrically coupled systems if noise asymmetry is
present. Thus noise asymmetry is found to complement the effect
of coupling asymmetry when the latter is absent.

We therefore conclude that, in modeling real systems where
synchronization arises, explicit consideration should be given
to the effect of possible asymmetries in coupling, noise, and
phase.

\section*{Acknowledgments}
The study was supported by the EC FP6 NEST-Pathfinder project
BRACCIA and in part by the Slovenian Research Agency and the
DST-Ramanna Fellowship of Prof. M. Lakshmanan, Government of
India.

\end{document}